\documentclass{elsart1p}
\usepackage{graphicx}
\usepackage{floatflt}
\usepackage{amssymb}
\usepackage{url}

\newcommand{\snn}{\sqrt{s_{NN}}}

\newcommand{\epem}{e^+e^-}

\newcommand{\np}{N_{part}}

\begin{document}
\begin{frontmatter}
%
%
%
\title{Soft Physics from RHIC to the LHC}
%
%
\author{Peter Steinberg}
\address{Brookhaven National Laboratory\\Upton, NY 11973\\USA}
\begin{abstract}
The RHIC program was intended to identify and study the quark-gluon
plasma formed in the collision of heavy nuclei.  The discovery of the
``perfect liquid'' is an essential step towards the understanding of
the medium formed in these collisions.  Much of data relevant to this
was provided by the study of ``soft'' observables, which involve many 
particles of low momentum produced in nearly every event, rather than 
high momentum particles produced in rare events.  The main results related to
soft physics at RHIC are discussed, as well as their implications for
the physics of the LHC heavy ion program.
\end{abstract}
\end{frontmatter}
%
\section{RHIC physics in a nutshell: The ``perfect liquid''}

\begin{figure}[t]
\begin{center}
\raisebox{3mm}{\includegraphics[height=42mm]{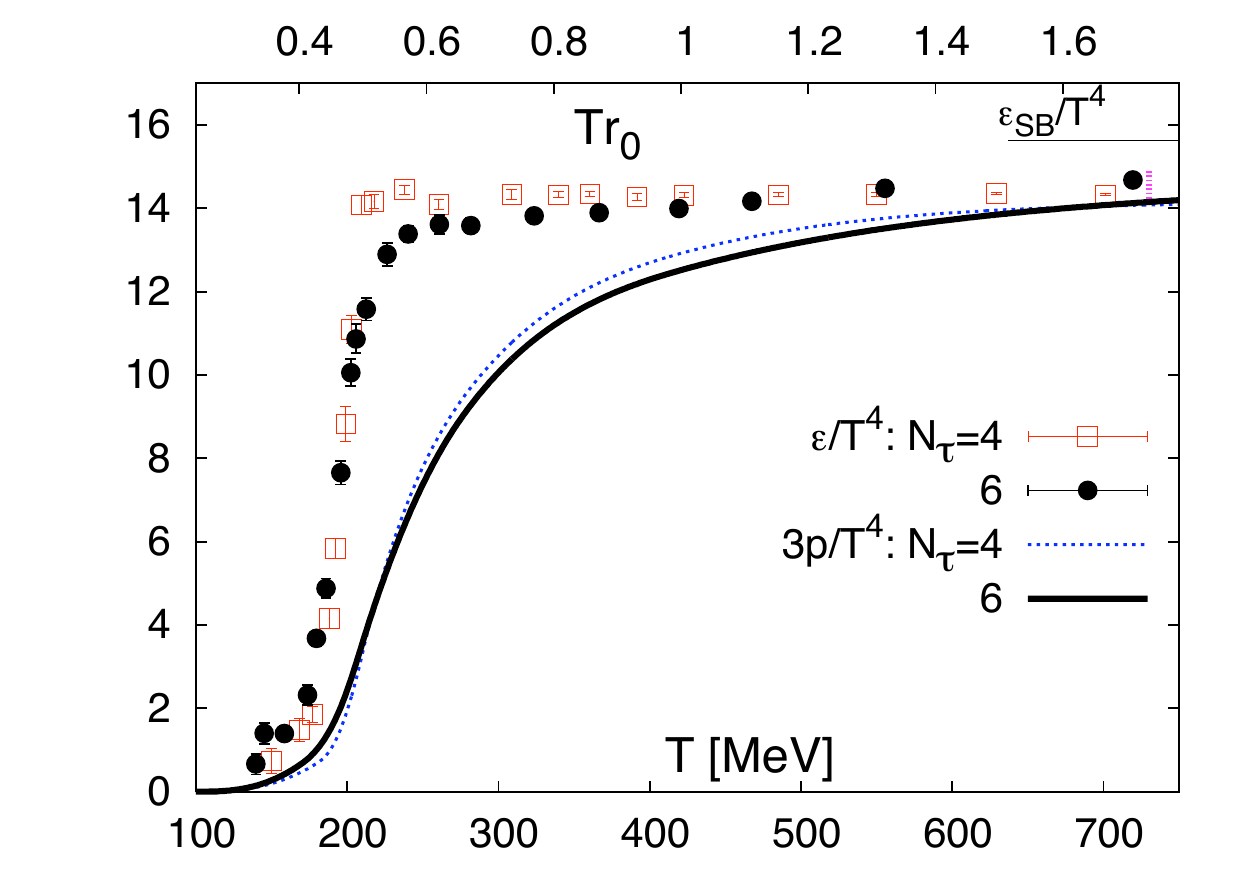}}
\includegraphics[height=47mm]{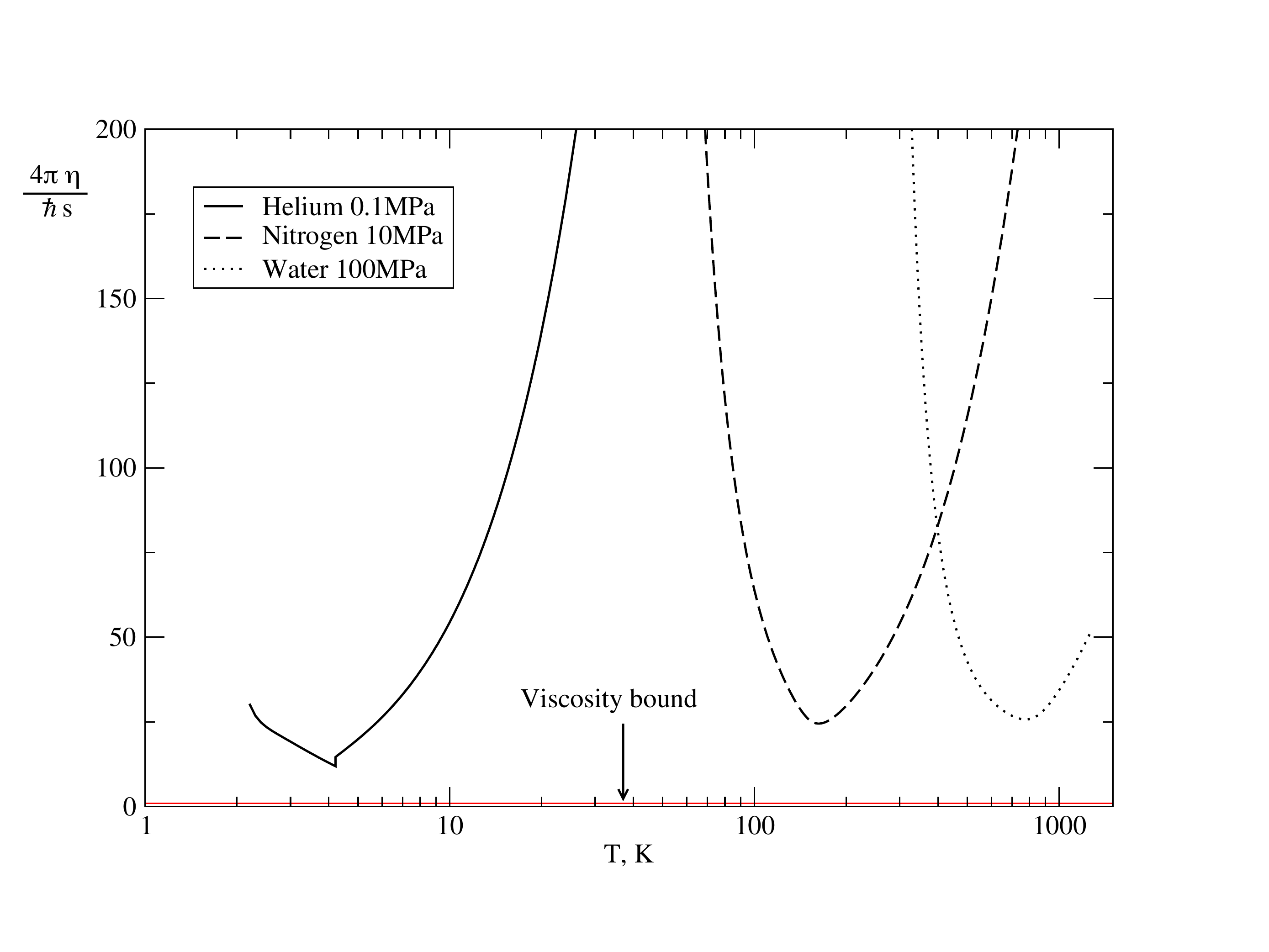}
\caption{
(left) Lattice QCD calculations showing the transition near $T_c = 170$ MeV from Ref.~\cite{Cheng:2007jq}.
(right) Viscosity scaled by entropy density as a function of temperature for three common liquids ($He$, $N$, $H_2 O$), normalized such that the viscosity bound is at unity.
\label{fig:QCD}}
\end{center}
\end{figure}

The Relativistic Heavy Ion Collider (RHIC) at Brookhaven National Laboratory turned on 
in 2000 and has taken data over a wide range of energies (19.6 to 200 GeV per 
nucleon-nucleon collision) and collision systems (protons to gold).  
The expectation from lattice QCD calculations (an example shown in the left panel of 
Fig.~\ref{fig:QCD}, from Ref.~\cite{Cheng:2007jq}) 
was that colliding nuclei at relativistic energies would form a hot, dense system
where the degrees of freedom would no longer be
the hadrons measured in the final state, but rather their quark and gluon constituents.
However, the data collected
by the four experiments at RHIC -- two large, and two small, all covering both 
unique and overlapping ranges in phase space -- arrived at the surprising conclusion
that the system formed in the collisions was a ``perfect liquid''~\cite{PL}.  
This is a non-trivial observation,
given that asymptotic freedom implied that
interactions between the quarks and gluons should become weaker at higher 
energies~\cite{Gross:1973ju}.  
Instead, the system appears to flow collectively as if the interactions between the relevant 
degrees of freedom are exceedingly strong.  The latter property can be 
quantified by the viscosity, which is large for an non-interacting gas, and zero at
infinite coupling.  Early estimates of the viscosity of the
medium formed at RHIC (discussed below) suggest that it is within a factor of 2 from the 
so-called ``viscosity bound'' predicted using the AdS/CFT 
correspondence~\cite{Kovtun:2004de}, shown compared to normal fluids in the right panel of Fig.~\ref{fig:QCD}.
This may well be the first prediction from string theory to be validated by experiment,
a major development for both heavy ion physics and string theory.

This work will discuss the most important results of ``soft'' physics at RHIC,
which involve the characteristics of multiparticle production at low momentum, rather
than the rare production of high momentum particles.  
Soft physics is of great practical value, as it is available to
experiments right after a machine starts.  And while it is not
yet amenable to perturbative approaches, it provides a handle on 
a sector of QCD of great interest for understanding the medium
produced in RHIC collisions.

\section{Evidence for the perfect liquid from soft observables}
\begin{figure}
\begin{center}
\raisebox{2mm}{\includegraphics[height=45mm]{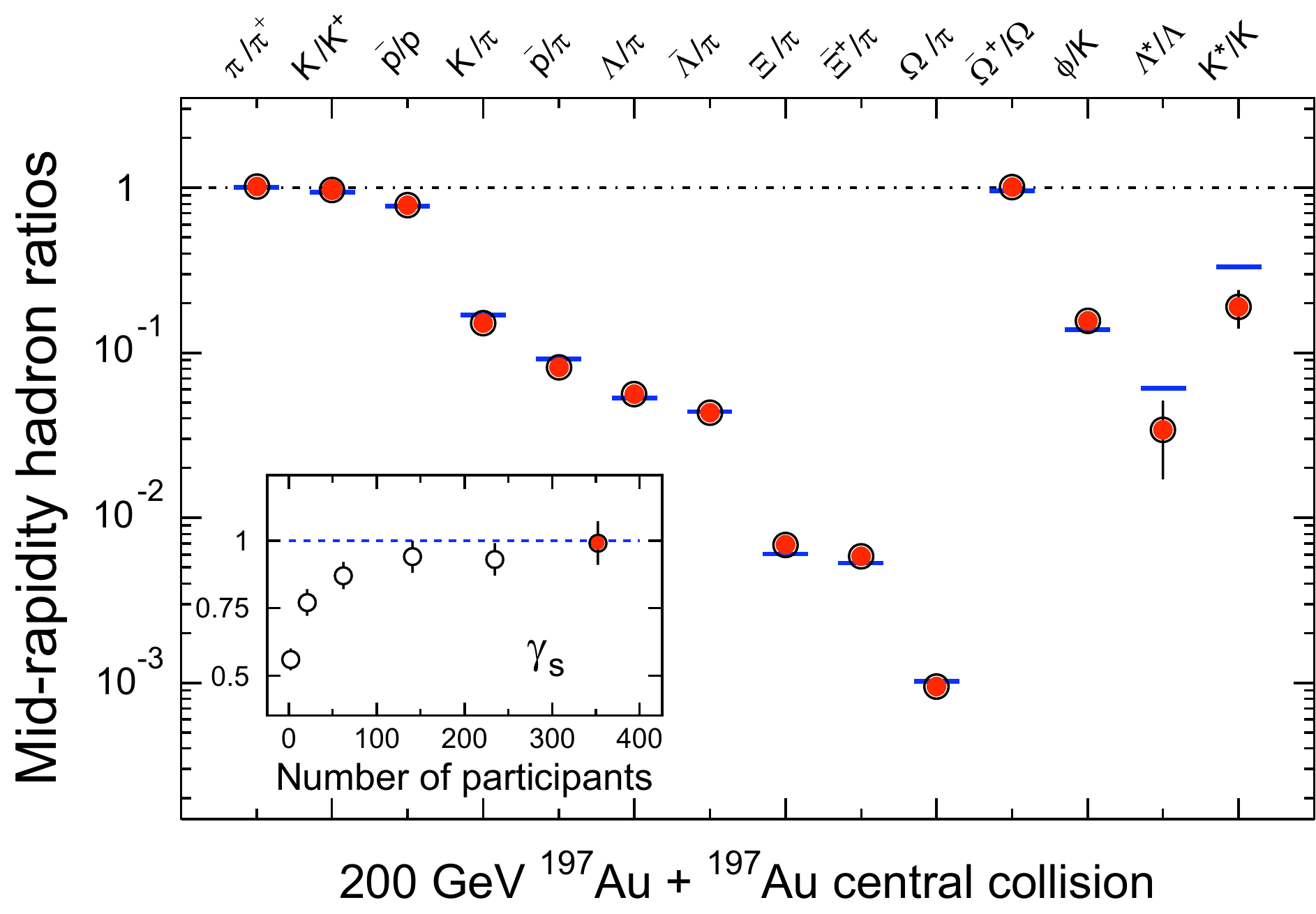}}
\includegraphics[height=45mm]{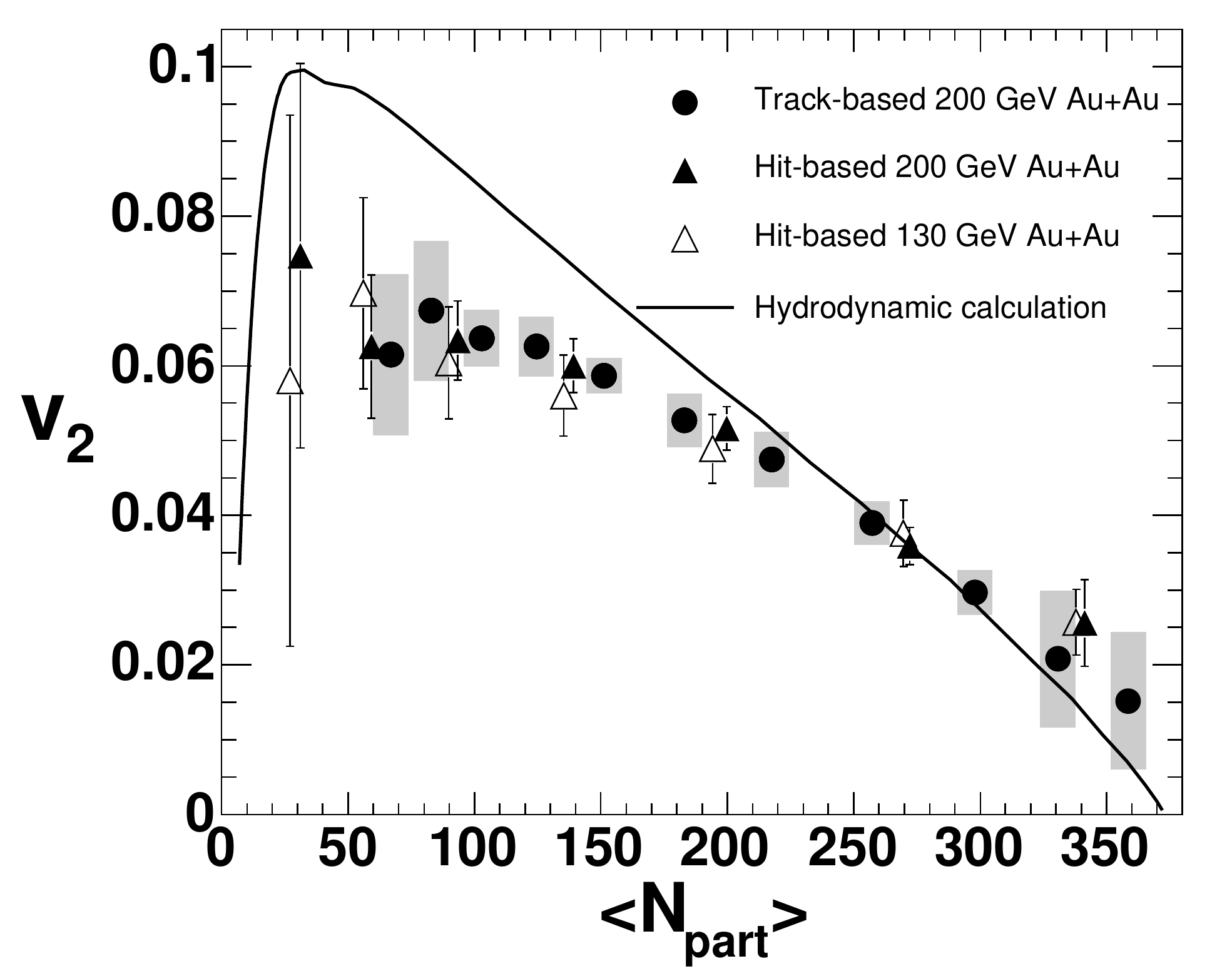}
\caption{
(left) STAR data on hadron ratios fit to a thermal model, from
Ref.~\cite{Adams:2005dq}.
(right) The elliptic flow parameter $v_2$ vs. $\np$ for $\snn = 200$ GeV
from PHOBOS compared with a hydrodynamical model.
\label{fig:thermalmodel}}
\end{center}
\end{figure}

The discovery of the perfect fluid at RHIC was established by a series of 
inferences from the final state back in time to the initial
state.  The spectrum of final state pions is manifestly blackbody in shape,
as one can see in distributions of identified 
particles (e.g. Ref.~\cite{Back:2006tt}).
Integrating the blackbody spectrum for the different hadronic mass states
(with proper Bose and Fermi distributions), and taking ratios
to factor out the emission volume, gives ``thermal model''~\cite{Cleymans:1992zc} 
predictions for ratios of total
yields of different hadron states.  Comparison to measurements, an 
example of which is shown in Fig.~\ref{fig:thermalmodel} from the STAR experiment~\cite{Adams:2005dq}
, shows an
excellent agreement and requires only three parameters: a chemical
freezeout temperature ($T_{ch}$), a baryochemical potential ($\mu_B$) reflecting
the excess of baryons in the initial state, and a strangeness suppression
factor (which is found to be nearly unity at RHIC).
These data suggest that the system was thermalized at the time of freezeout and was
thus at least $T = 170$ MeV during its evolution.

The success of describing the data from such a wide variety of hadron states
validates the approach begun by Rolf Hagedorn in the 1960's, who observed 
the thermal slopes in hadron momentum spectra and postulated that they
were emitted from a thermalized 
state~\cite{Hagedorn:1965st}.  
The observation of the same temperature
in all systems led him to also propose a ``limiting temperature'' arising 
from an exponentially rising mass spectrum ($dN/dM \propto \exp(M/T_H)$).  
The rising mass spectrum 
was a prediction borne out by forty years of subsequent measurements, making
surpassing the Hagedorn temperature $T_H \sim 170$ MeV that much more interesting
for heavy ion experiments forming hot and dense matter.  Lattice
QCD calculations appear to have no limiting temperature, reflecting the fact
that they find a discontinuity in the number of degrees of freedom, e.g. as 
quantified by $\epsilon/T^4$ shown as a function of T in the right panel of
Fig.\ref{fig:QCD}, at a temperature quite similar to $T_H$~\cite{Cheng:2007jq}.

Beyond being thermal, heavy ion collisions show a collective flow characteristic of a liquid.
At all energies and system sizes, it has been observed
that the event-by-event angular distributions are not isotropic
in azimuthal angle.  An ``event plane'' can be estimated from the 
produced particles, defined as angle $\Psi_R$ of 
the short principal axis of the particle angles.
The azimuthal distribution relative to this event plane is found
to show a strong $\cos(2[\phi-\Psi_R])$ dependence.  This is
especially pronounced in more peripheral (i.e.\ lower multiplicity)
collisions, where the overlap of the nuclei is shaped like an
almond, relative to central (i.e.\ higher multiplicity) collisions
where the overlap is essentially isotropic.  This leads to
a characterization of the event-by-event angular distributions in
terms of its Fourier coefficients~\cite{Voloshin:1994mz}
$\frac{dN}{d\phi} = 1+2v_1 \cos(\phi-\Psi_R) + 2v_2 \cos(2[\phi-\Psi_R]) + \cdots$.
PHOBOS data on $v_2$ is shown in Fig.~\ref{fig:thermalmodel} and compared to
a hydrodynamical calculation tuned to the most central data and extrapolated to
larger impact parameters using a Glauber 
model~\cite{Back:2004mh}.

\begin{figure}[t]
\begin{center}
\includegraphics[height=50mm]{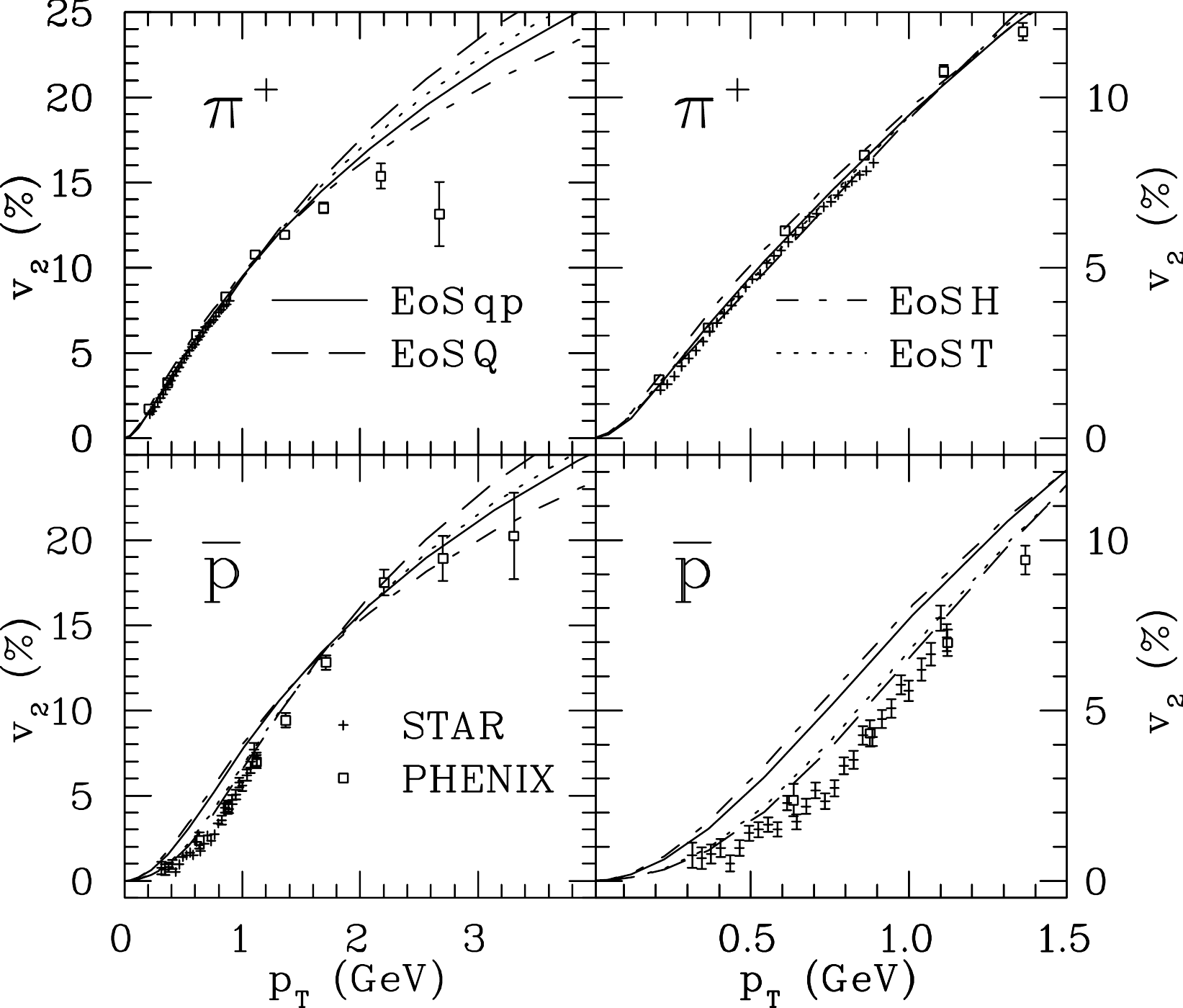}
\includegraphics[height=55mm]{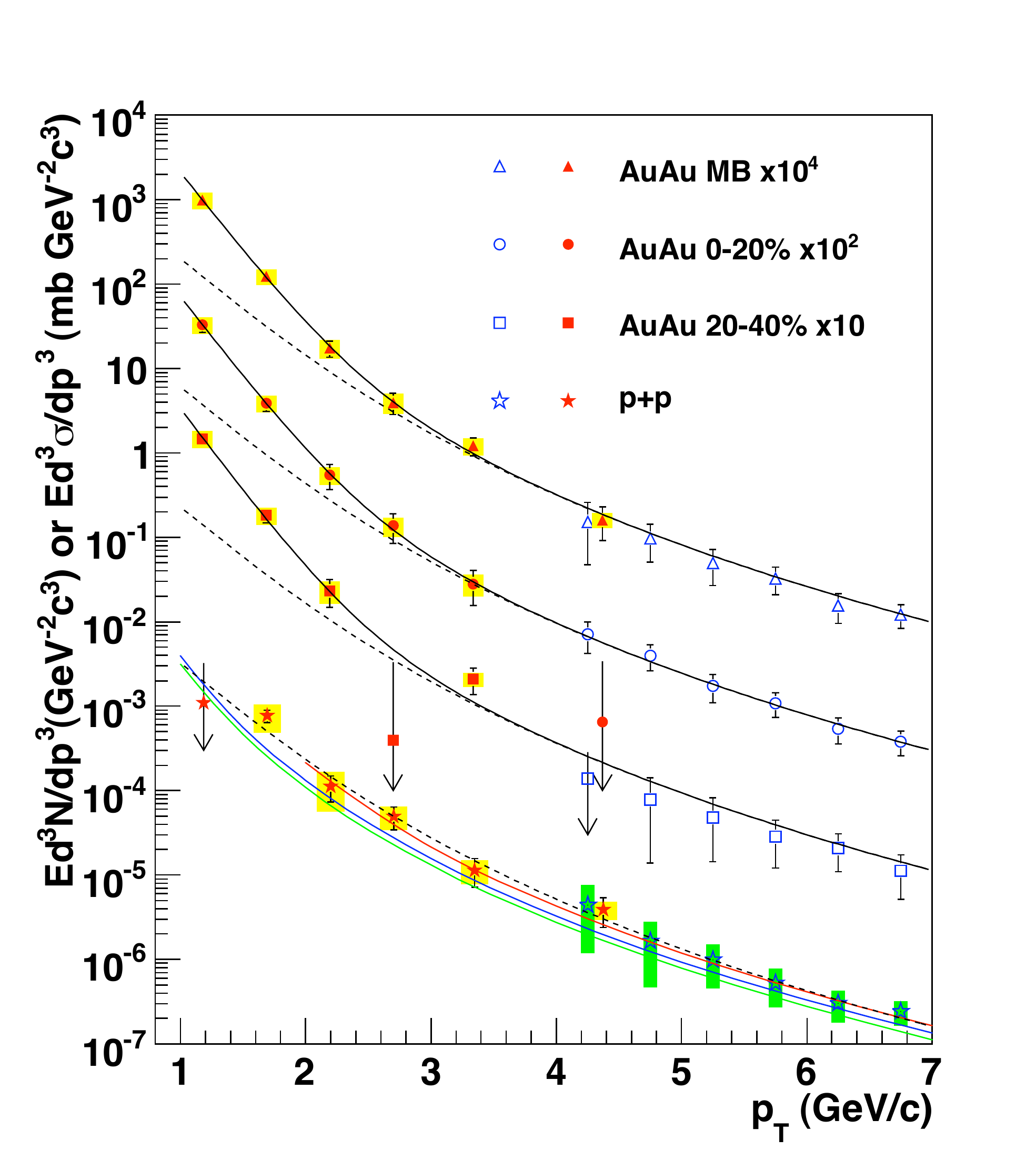}
\caption{
\label{fig:v2piap5}
(left) Effect of different equations of state on pion and proton flow, from Ref.~\cite{Huovinen:2005gy}.
(right) Direct photon spectra measured by PHENIX for p+p collisions and for heavy ion collisions over a range of impact parameter.
}
\end{center}
\end{figure}

Comparing the relevant energy and space-time scales implied by the
success of the hydrodynamical models, the matter at RHIC is formed
under quite extreme conditions.  The formation time needed for the
the hydrodynamic calculations is $\tau_0 = 0.6$ fm/c, or 
approximately 
2 yoctoseconds ($10^{-24}$)~\cite{Kolb:2000fha}.  
This number is far smaller than the
time taken a massless particle to traverse the radius of a hadron
($\tau \sim 1$ fm/c)~\cite{Bjorken:1982qr}.
The same calculations determine that the energy density needed to match the data is around
$\epsilon \sim 30$ GeV/fm$^3$, 
about 60 times the density of a nucleon in its rest frame, $\epsilon_N \sim 500$ MeV/fm$^3$.
It should be noted that these estimates do not preclude even higher energy densities at
even earlier times.

The agreement of hydrodynamic models with RHIC collisions is an important, 
but surprising, development.
Hydrodynamics is not merely ``another'' model, but describes
a medium with particular properties, difficult to generate by the rescattering of 
partons or hadrons.
It is a locally equilibrated system, with a well-defined temperature at
any given space-time location.  The agreement with data, particularly 
the detailed flow of heavy particles, seems to require a ``lattice-like''
equation of state, with a speed of sound that drops rapidly at the
hadronization temperature.  
With an energy density at least 60 times the energy density of a proton, the system resists
a description as interacting hadrons, as their wave functions would all be overlapping
and thus indistinguishable as individual particle states.  

This same energy
density implies a temperature about 2-3 times the critical temperature found on 
the lattice, well above the Hagedorn temperature.
Exceeding the Hagedorn temperature is a key piece of evidence required for any system 
claiming to have degrees of freedom comprised of quarks and gluons.
The recent PHENIX data, shown in the left panel of Fig.~\ref{fig:v2piap5}, show a spectrum
of direct virtual photons from heavy ion collisions, compared with a similar
sample from proton-proton collisions~\cite{:2008fqa}.  There is a notable excess in the
lowest $p_T$ bins which can be fit by a simple exponential of slope 
$T \sim 220$ MeV.  These data do not fully establish the temperature
of the system at the very early times, as the measured spectrum is a
convolution of spectra emitted during the whole collision evolution.  
Still, they directly establish that the emitted photons were hotter than 
the Hagedorn temperature at some point in the collision evolution.  
This suggests that the medium may well be the
QGP, although this result {\it per se}
does not point directly to quark and gluon degrees of freedom.
Interestingly,
comparisons of data for pions and protons shown in the right panel of 
Fig.~\ref{fig:v2piap5} require a sharper transition than
actually observed in lattice calculations~\cite{Huovinen:2005gy}.


\section{The edge of liquidity: Can we turn off the perfect liquid?}

The detailed characterization of the medium formed at RHIC requires a systematic 
variation of the initial conditions to measure its effects on the
final state particles.  Such scans have been performed in other systems,
e.g. ultra-cold atomic gases~\cite{Thomas}.  Since soft physics involves
the bulk production of particles from a thermalized system undergoing collective
flow, one would want to vary the temperature, density and material properties
in such a way to quantitatively extract the equation of state.  Unfortunately,
the only tools available for heavy ion physics are colliding ion beams,
for which one can only control the beam energy and nuclear size.  
It also possible to vary the centrality of the collision, which affects the shape of the
overlap region.

The most fundamental question addressed in all of these experimental studies is if
the system is sufficiently thermalized in the initial state, such that it can
properly be called a state of matter.  Once that is established,
a whole series of questions arises, each of which addresses fundamental problems in 
strong interaction physics.  
\begin{itemize}
\item Does the system thermalize everywhere, or just in limited regions of phase space?
\item What are the conditions (size, energy) for thermalization?
\item How rapidly is thermalization achieved?
\item What degrees of freedom reach thermalization (partons, hadrons, or something else)?
\end{itemize}
If it is assumed that the top-energy, most-central RHIC collisions are in fact thermalized
(as suggested by their agreement with hydrodynamical models),
it should be possible to vary the experimental parameters and ``turn off'' 
the creation of quark gluon plasma. 

\begin{figure}[t]
\begin{center}
\raisebox{1mm}{\includegraphics[height=40mm]{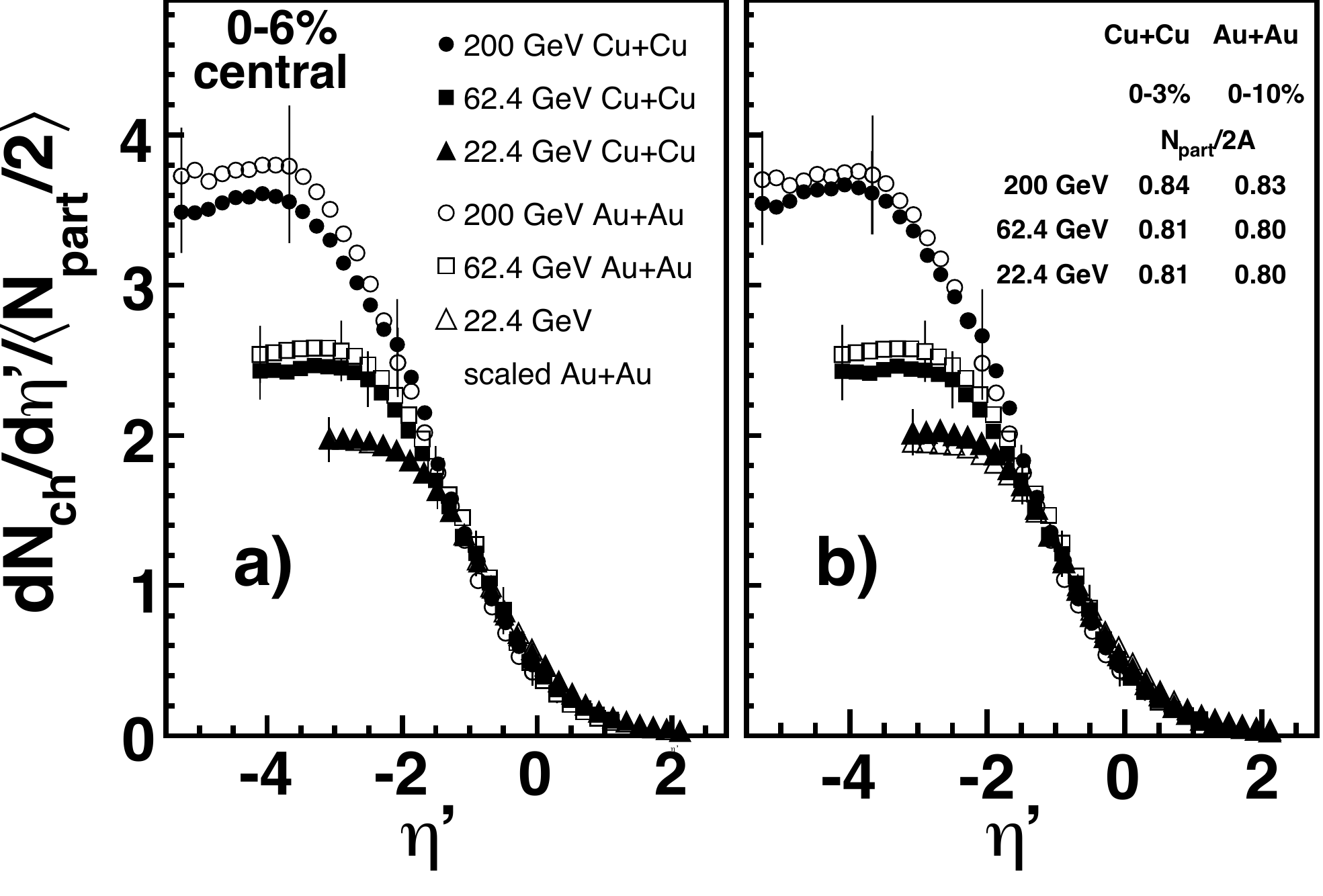}}
\includegraphics[height=45mm]{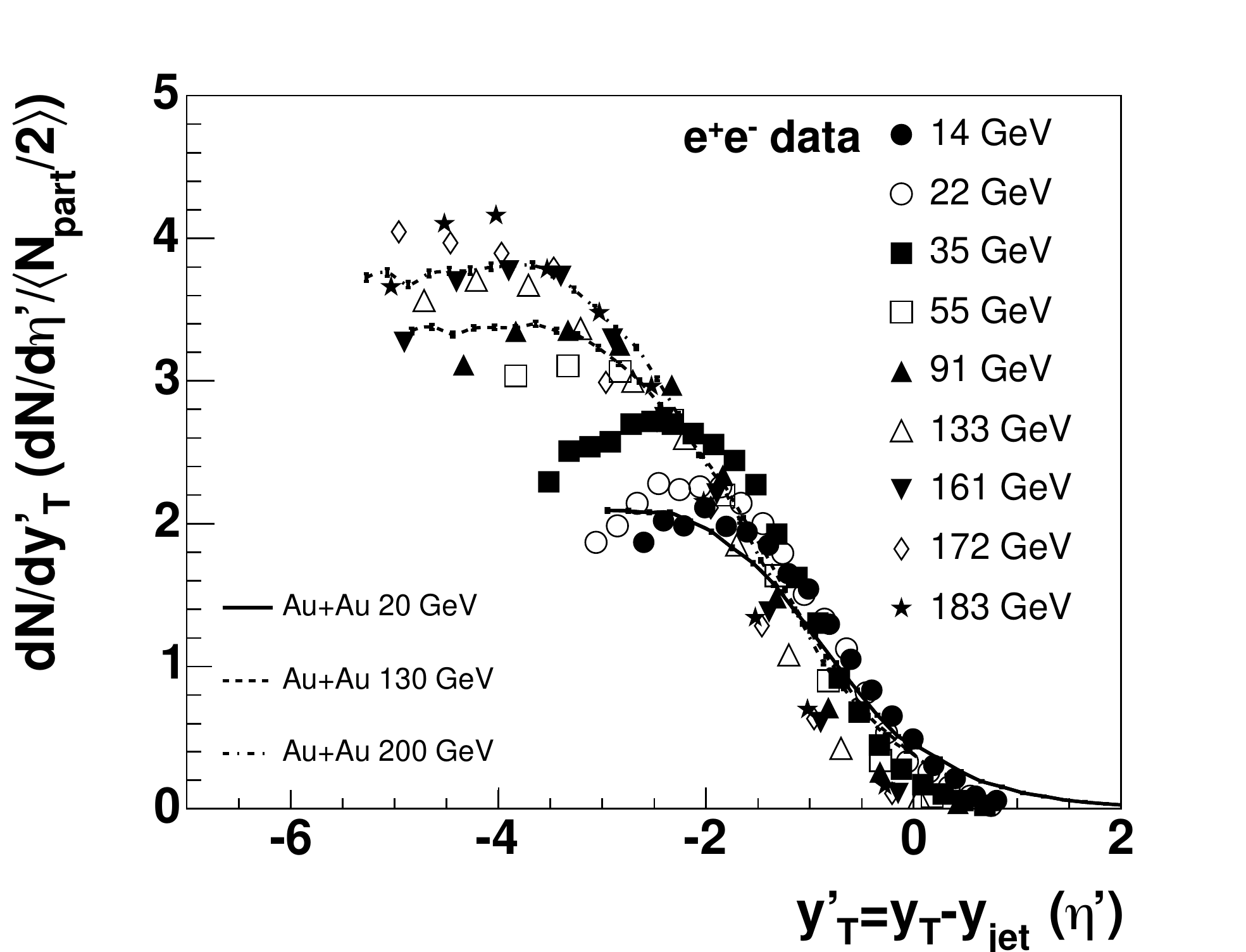}
\caption{
\label{fig:phobos-limfrag}
(left) Extended longitudinal scaling observed for inclusive charged particle distributions in Au+Au collisions over a wide range of energies~\cite{:2007we}.
(right) Extended longitudinal scaling seen in $\epem$ reactions compared to heavy ion data (figure from Ref.~\cite{Steinberg:2004wx}).
}
\end{center}
\end{figure}

\begin{floatingfigure}[r]{55mm}
\begin{center}
\includegraphics[width=45mm]{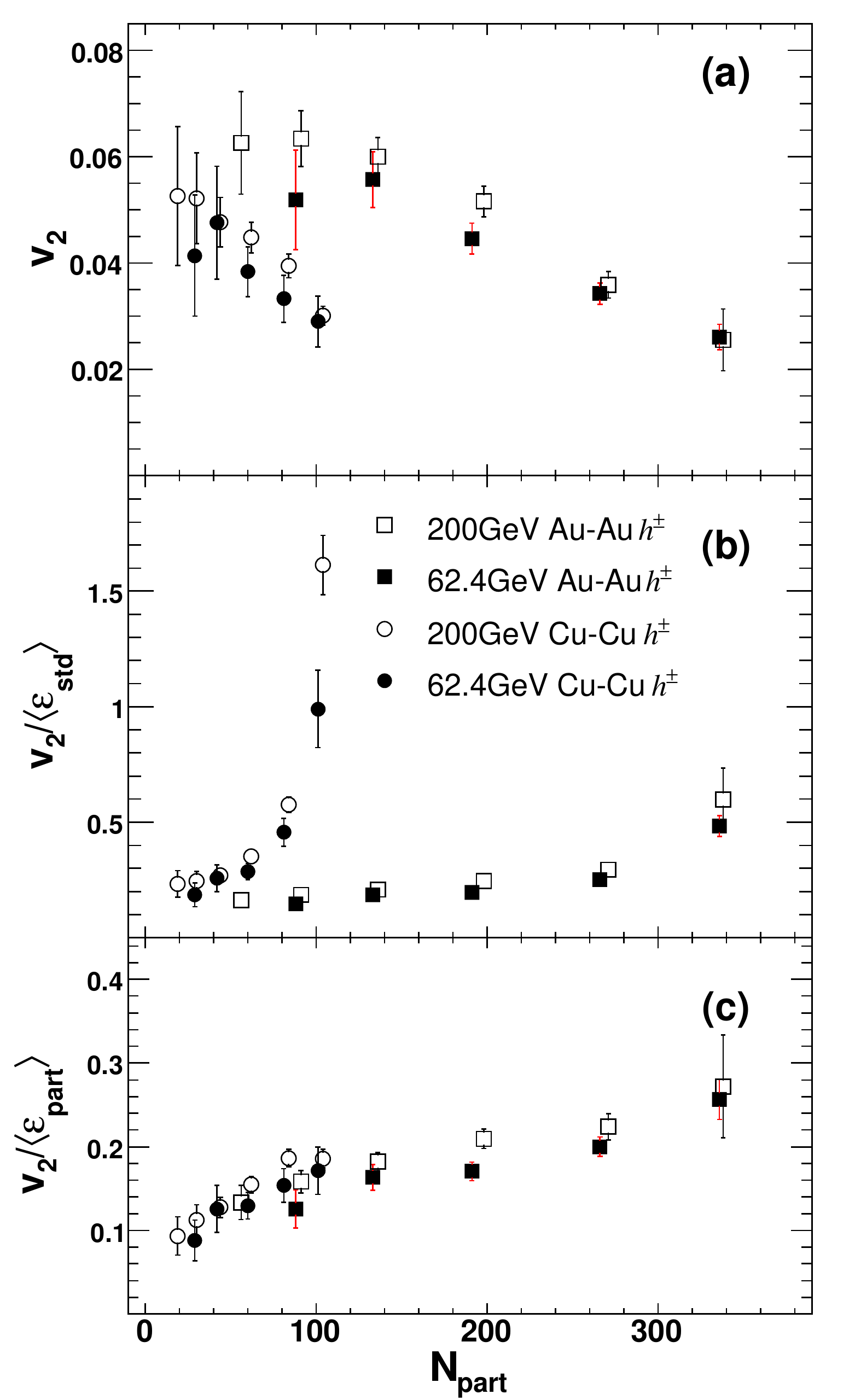}
\caption{
\label{fig:eccpart}
$v_2$, $v_2/\epsilon_{std}$, and
$v_2/\epsilon_{part}$ from PHOBOS data~\cite{Alver:2006wh}.
}
\end{center}
\end{floatingfigure}

Lowering the beam energy would seem to be a natural way to create a system 
that does not form a quark gluon plasma.
However, it is essential to check if this leads to 
particle production different in some essential way than at higher energies.
The inclusive charged particle multiplicity is a useful measurement for this, as it 
is a ``global'' observable, which integrates over time, space, and various degrees of freedom.
The number of particles emitted essentially
counts the degrees of freedom available to the system, which should be linear with the
produced entropy in a fully thermalized 
system~\cite{Belenkij:1956cd}.  
One empirically observed feature that should shed light on this is 
``extended longitudinal scaling''~\cite{Back:2002wb}.
It has been observed in proton-proton collisions that the 
pseudorapidity density ($dN_{ch}/d\eta$) of
inclusive charged particle production is energy-independent 
when viewed in a frame where one or other of the incoming particles is at 
rest~\cite{Alner:1986xu}.
This is done by using the kinematic variable $\eta^{\prime} = \eta - y_{beam}$, 
where $y_{beam}$ is the rapidity of one of the beams.  
Results for $dN/d\eta^{\prime}$ are shown for Au+Au collisions at 
four RHIC energies~\cite{:2007we} in the left panel of Fig.~\ref{fig:phobos-limfrag}, 
where longitudinal scaling is clearly observed.  
The persistence of this scaling over a factor of ten in energy suggests that no major
changes in the particle production occurs over this range.  
It also suggests that physics at $\eta=0$ is not 
obviously from a different origin than physics in the forward direction.

What is surprising is that the same phenomena is common to all systems, 
from heavy ions to proton-proton, and even to $\epem$ annihilation.
The latter system is shown in the right panel of Fig.~\ref{fig:phobos-limfrag}, which shows
that the phenomenon is also not dependent on the overall system size~\cite{Back:2004je}.  
It also shows an overlay of three Au+Au energies~\cite{Steinberg:2004wx},
showing that the overall magnitude of multiplicity in
heavy ion collisions and $\epem$ reactions is essentially the same, provided the Au+Au
data is scaled down by the number of participant pairs ($\np/2$).
While this similarity between heavy ions and $\epem$ has been reported 
for several years~\cite{Back:2006yw}, it has not been fully explained.

PHOBOS data in Ref.~\cite{Back:2004zg} show that the elliptic flow 
parameter $v_2$ also obeys extended longitudinal scaling.
This is surprising if the initial conditions at midrapidity 
are completely different to that at forward rapidities,
but not if the initial entropy density and geometry are the 
primary determinants of elliptic flow, as should be the case
if the viscosity is small.  It has been shown by various experiments
that the elliptic flow, normalized by the initial eccentricity, scales with the
transverse particle density $dN/dy/S$, where $S$ is the area overlap of the initial 
collision~\cite{Adler:2002pu}.
However, this scaling breaks down when comparing Au+Au and Cu+Cu reactions 
if the ``standard'' eccentricity is used:
$\epsilon_{std} = (\sigma^2_Y - \sigma^2_X)/(\sigma^2_Y + \sigma^2_X)$, 
where $\sigma^2_X$($\sigma^2_Y$) is the variance in the direction along 
(perpendicular to) the reaction plane.
Fig.\ref{fig:eccpart} shows that scaling is restored by using an eccentricity defined by the
distribution of the participants themselves~\cite{Alver:2006wh}:
$\epsilon_{part} = (\sqrt{ (\sigma^2_Y-\sigma^2_X)^2+4(\sigma^2_{XY})^2 })/(\sigma^2_X + \sigma^2_Y)$.
This suggests that the geometrical configuration
of the participants is ``frozen in'' immediately,
consistent with the previous estimates of $\tau_0$ 
or perhaps even shorter times.

\begin{figure}[t]
\begin{center}
\raisebox{1mm}{\includegraphics[width=50mm]{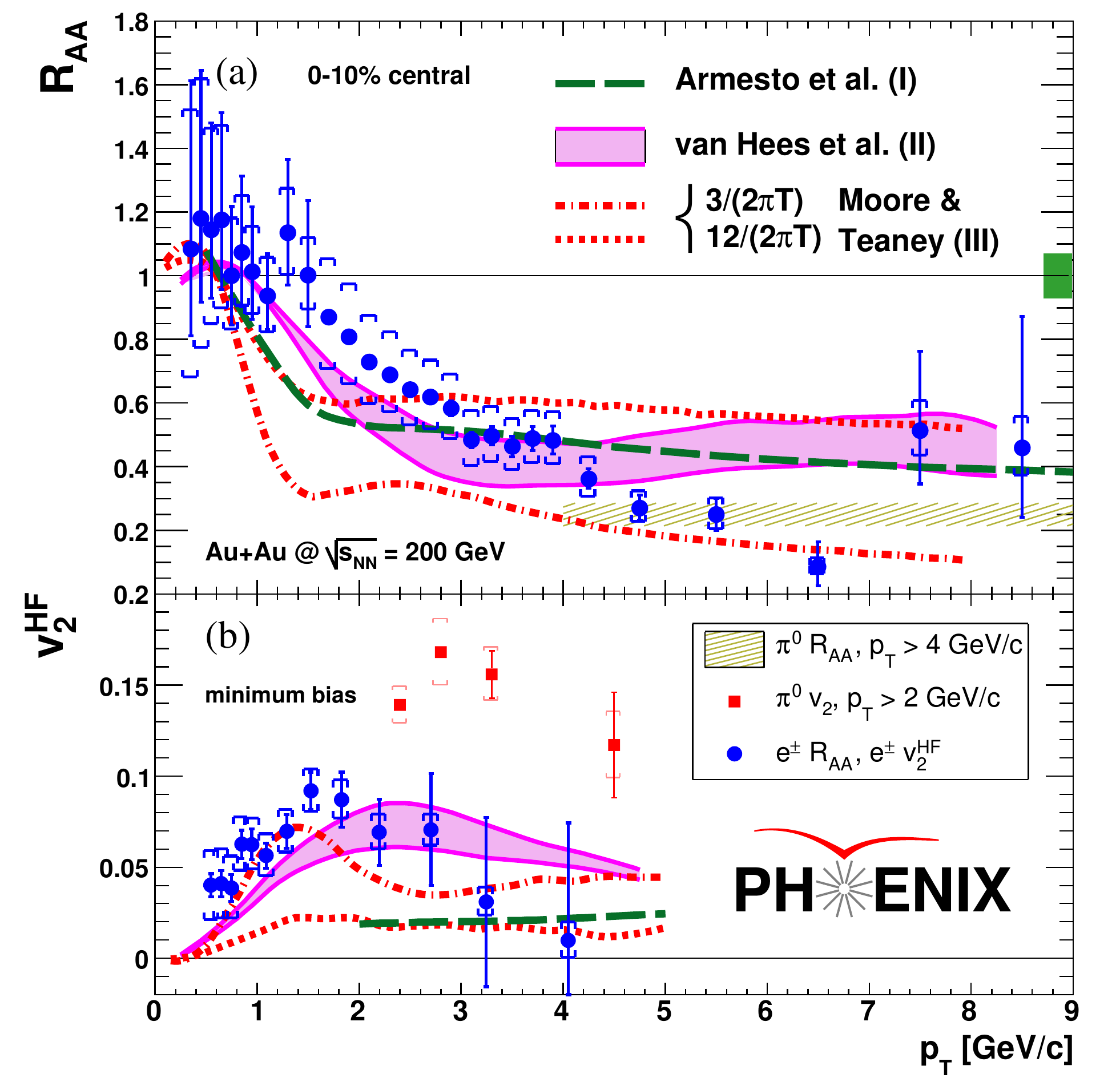}}
\includegraphics[width=68mm]{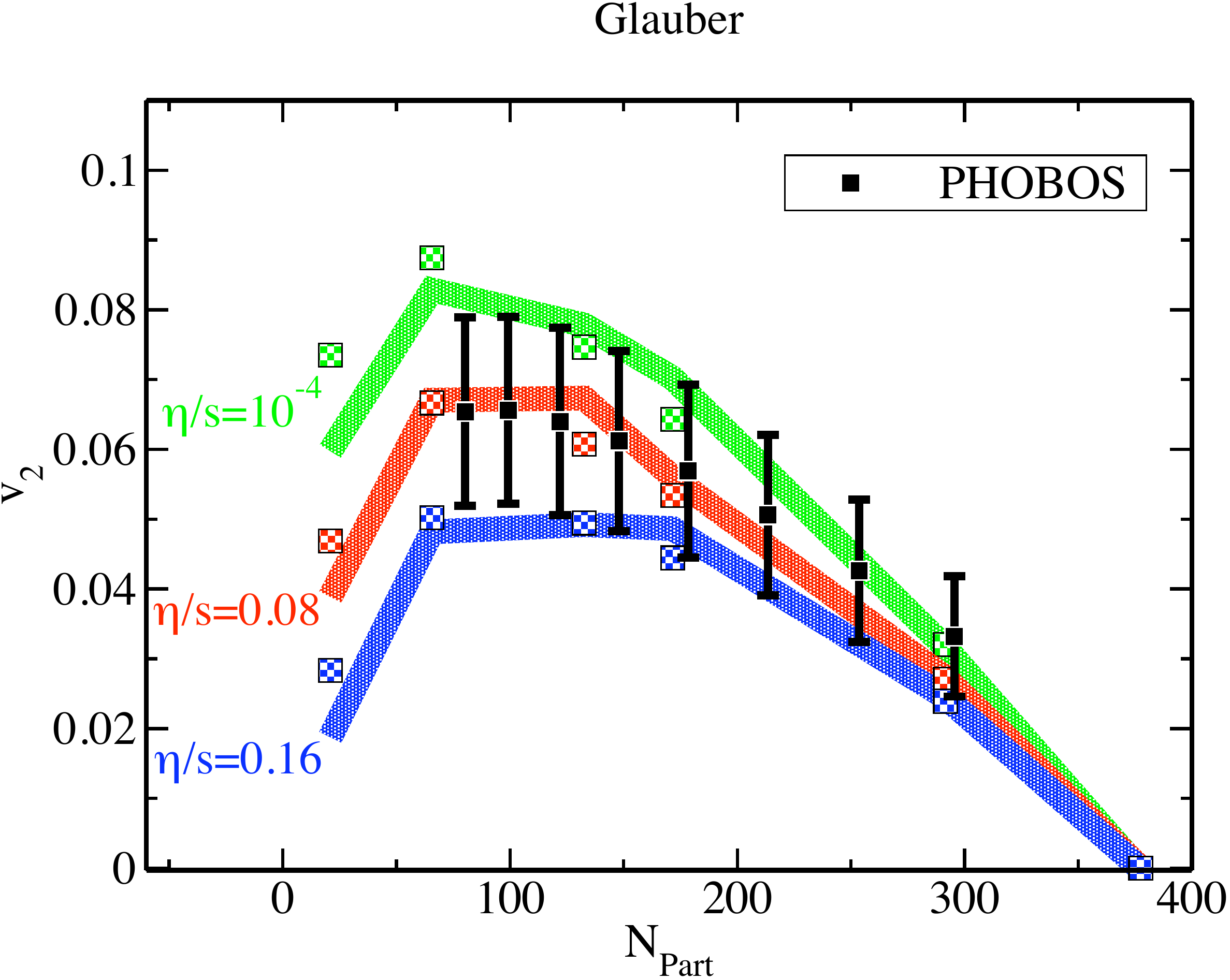}
\caption{
\label{fig:phenix_charm_fig3}
(left) PHENIX data Ref.~\cite{Adare:2006nq} on charm suppression and elliptic flow
for Au+Au collisions.
(right) PHOBOS data compared to a set of calculations using the viscous hydro code from Ref.~\cite{Luzum:2008cw}.
}
\end{center}
\end{figure}

The predictions from AdS/CFT for shear viscosity~\cite{Kovtun:2004de}
have made measurements of observables
sensitive to viscous effects a high priority.  
Currently, there are two approaches 
for making quantitative estimates of $\eta/s$ based on experimental data.  
One is based on direct measurement of non-equilibrium
processes, e.g. the behavior of charm quarks which are produced in the earliest stages.  
Model calculations are used to relate the
observed magnitude of $v_2$ with simultaneous measurements of suppression as a function of $p_T$.
PHENIX results~\cite{Adare:2006nq}, 
shown in Fig.\ref{fig:phenix_charm_fig3}, find good agreement between the
data and models assuming that $\eta/s \sim (1-2)/4\pi$.  The other method is to systematically
include viscous corrections into the hydrodynamic evolution, a technically challenging
task which is being addressed by several competing teams~\cite{TECHQM}.  
An example of this from Ref.\cite{Luzum:2008cw}
is shown in the
right panel of Fig.~\ref{fig:phenix_charm_fig3}, and compared to PHOBOS data.  The main 
uncertainty in these calculations at present is the initial state (e.g. whether the 
energy is distributed according to Glauber models or according to the Color Glass Condensate).
However, the effect on $\eta/s$ is only about a factor of two, which suggests that RHIC
is quite close to the viscosity bound.

\section{Degrees of freedom: What is the perfect liquid?}

If the near-perfect fluid description is relevant over
most of the evolution, it is essential to determine
the fluid constituents, and how they become a thermalized
collective state of matter.
Nothing in the data discussed so far uniquely identifies which
degrees of freedom are able to achieve this.
The most natural assumption would be that the early stages are
dominated by the dynamics of free quarks and gluons, or at
least the dynamics of quark and gluon fields that are studied
using Lattice QCD.  However, as shown above attempts to model the existing
data on $v_2$ vs. $p_T$ for identified particles find that 
a strong first-order phase transition needs to be put in 
by hand~\cite{Huovinen:2005gy}.
Existing lattice calculations, shown in Fig.~\ref{fig:v2piap5} 
do not provide sufficient ``softening'' of the equation of 
state to reproduce the heavier particles which are most sensitive 
to the speed of sound.


\begin{figure}[t]
\begin{center}
\includegraphics[width=60mm]{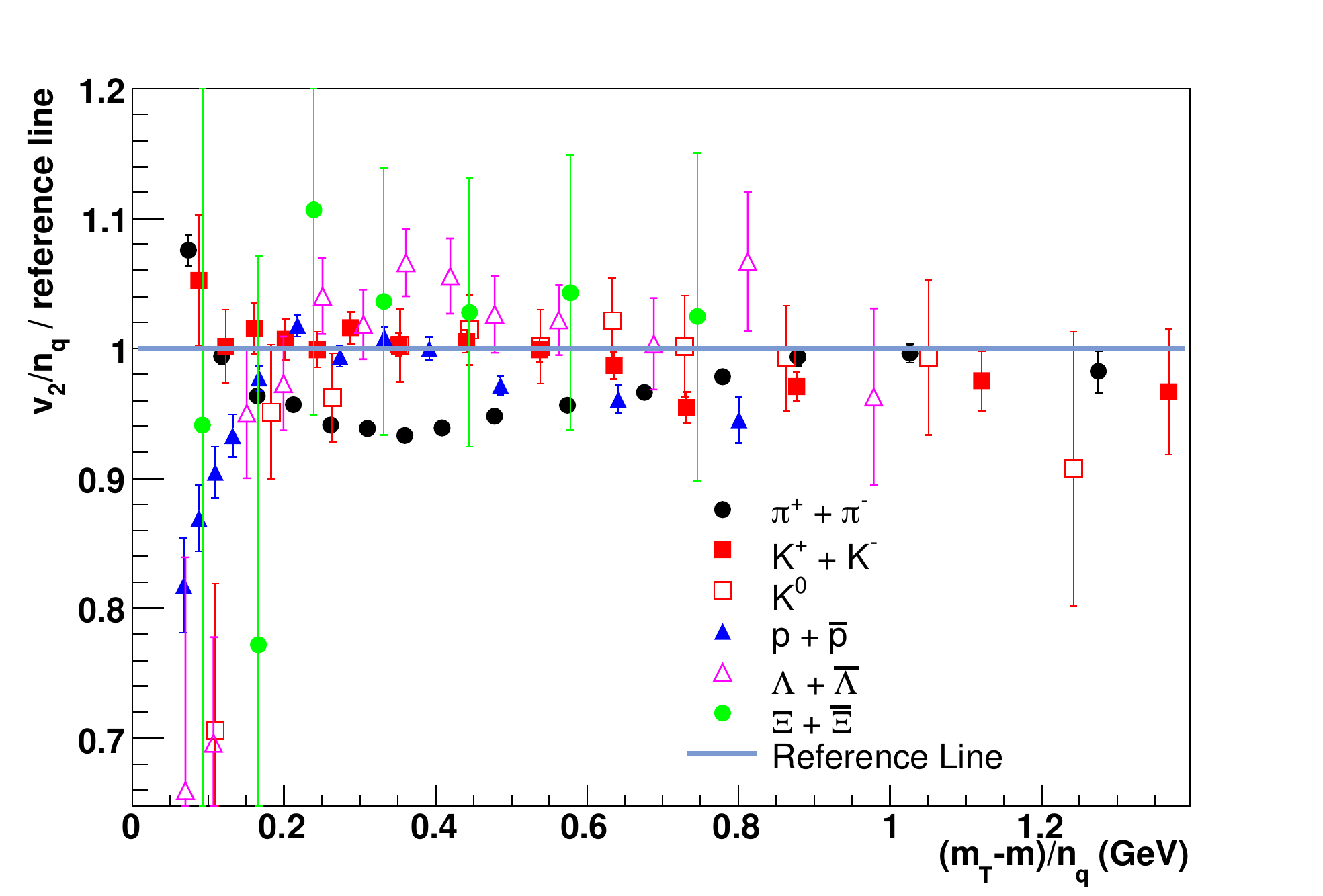}
\includegraphics[width=60mm]{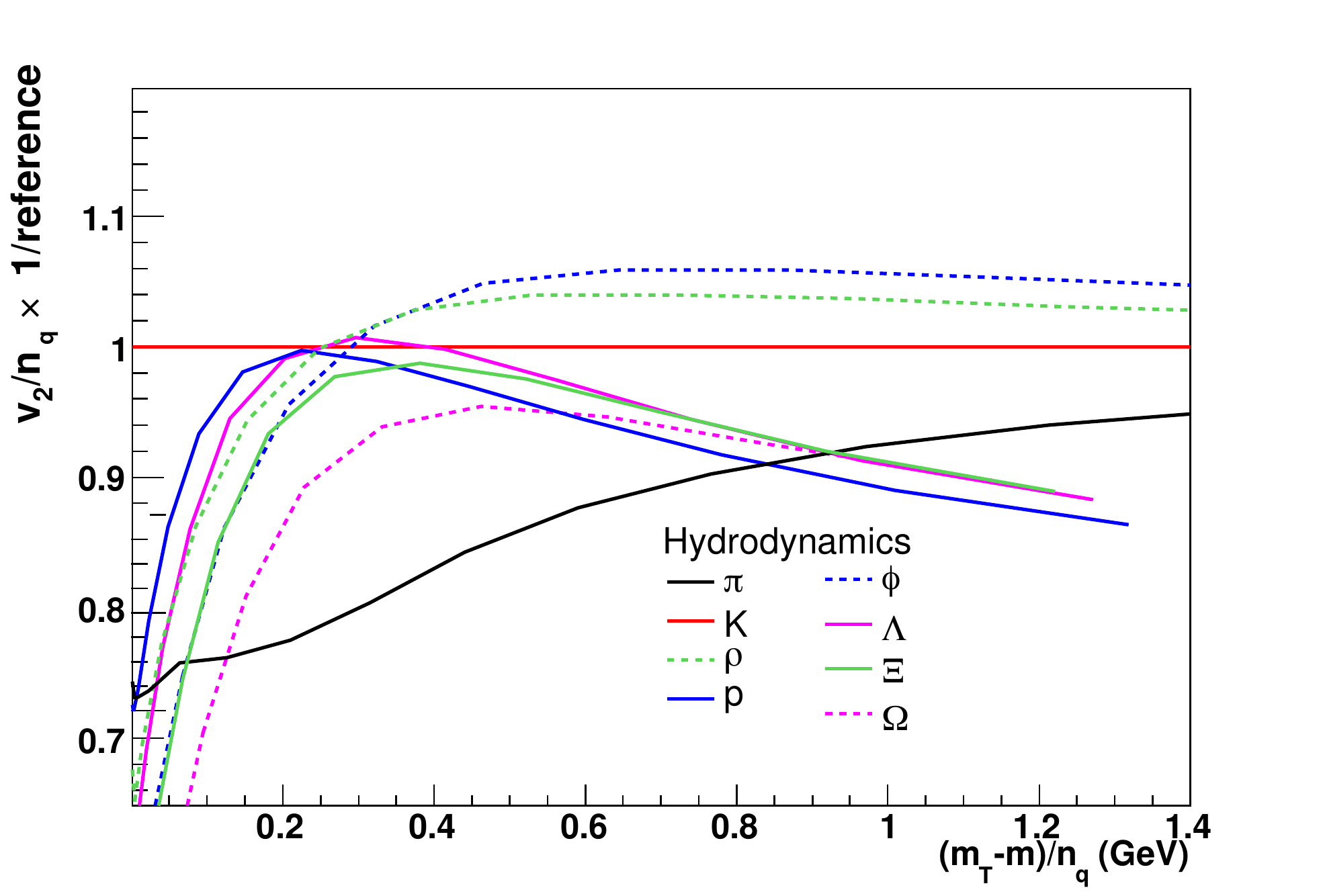}

\caption{
\label{fig:v2svsketsref}
(left) PHENIX data on $v_2/n_q$ (from Ref.~\cite{Adare:2006ti}, divided by a reference) vs. $KE_T/n_q$ for various hadrons
(right) the same quantity for a hydrodynamic calculation, both from Ref.~\cite{LindenLevy:2007gd}.
}
\end{center}
\end{figure}

Another way to look for quark degrees of freedom is via constituent quark ($n_q$) scaling.
The $v_2$ data have been studied for many different particles species,
which have different mass and quark content.  
PHENIX data~\cite{Adare:2006ti}
show that all of the available data on $v_2$ vs. $p_T$
lie near one another when plotted as a function of 
$v_2/n_q$ on the Y axis, where $n_q$ is the number of valence
quarks and anti-quarks in the hadron, and $KE_{T}/n_q$ on
the X axis, where $KE_{T} = m_T - m$ for each hadron of mass
$m$.  This suggests a scenario where freezeout occurs by the
recombination of constituent quarks, assumed to 
have a mass of $\sim m/n_q$ and the appropriate quantum numbers for
each hadron.  However, Ref.~\cite{LindenLevy:2007gd} argues that the presence of good
quasiparticles, where the width is small relative to the mass,
should induce viscous effects which could exceed the estimates of
viscosity using the data in Fig.~\ref{fig:phenix_charm_fig3}.
And yet, the $v_2/n_q$ data scaled by a reference derived from the kaon flow
(shown in the left panel Fig.~\ref{fig:v2svsketsref}) show that the flow data 
fit into the $n_q$ scaling pattern, while hydro calculations
(shown in the right panel) unsurprisingly do not~\cite{LindenLevy:2007gd}.  
These data imply that it remains an open question how to harmonize
the perfect fluid and constituent quark scenarios.  
If one tries to recapture the fluid behavior by allowing the 
large quasiparticle widths, it is not clear at which point the quasiparticles
no longer act as the relevant degrees of freedom.  

\section{The future: RHIC II and the LHC}

\begin{figure}[t]
\begin{center}

\raisebox{1mm}{\includegraphics[width=60mm]{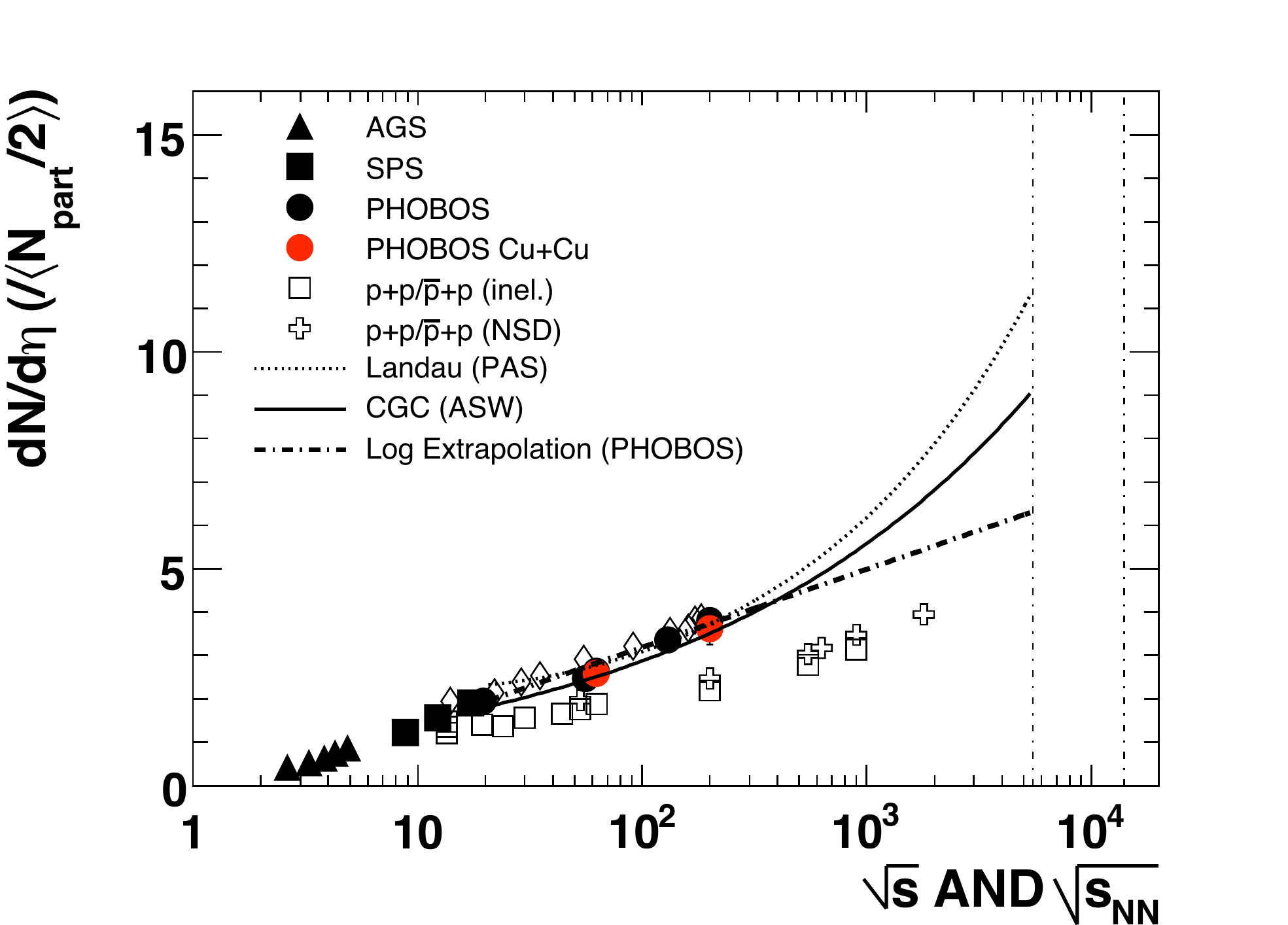}}
\includegraphics[width=60mm]{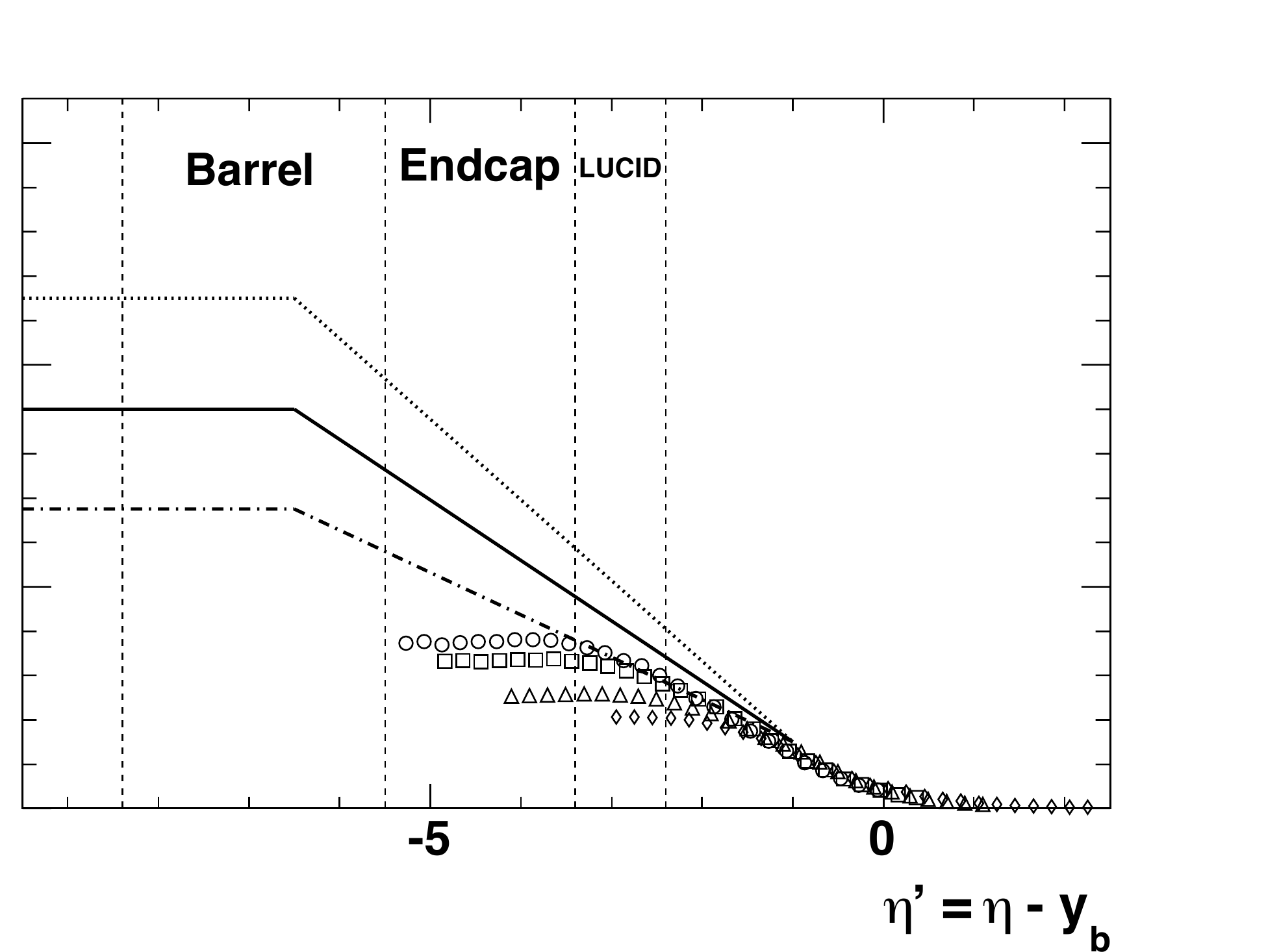}
\caption{
\label{fig:lhc}
(left) Data on mid-rapidity multiplicity in p+p and A+A, with predictions for the LHC energies, from Ref.~\cite{Steinberg:2008zz}.
(right) Coverage of LHC experiments in the forward direction, presented in the rest frame of one of the projectiles.
}
\end{center}
\end{figure}

The perfect liquid is now the paradigm that will be extensively tested by the next generation
of heavy ion experiments.  These will take place at an upgraded RHIC II,
with a factor of ten increase in luminosity, and at the upcoming LHC Pb+Pb
program, with a factor of nearly thirty increase in center-of-mass energy.
Probing the transport properties of the system with jets and heavy quarks will be the
focus of the next generation of RHIC experiments~\cite{RHICII}.  
New silicon detectors in PHENIX and STAR are being developed
to measure charmed particles by means of displaced decay
vertices.  
The LHC will explore both hard and soft physics in a new energy regime, with
the three large LHC experiments -- ALICE, ATLAS and CMS  -- all with active
heavy ion programs~\cite{Carminati:2004fp,Grau:2008ef,D'Enterria:2007xr}.  
Hard probes will be especially
powerful, due to the increased rate of jets, photons, and quarkonia expected
from perturbative QCD calculations.  

However, soft physics will also lead to great strides in our 
understanding of the strongly-coupled medium.  Furthermore, the results will 
come out quickly, and be able to immediately address the relevance of models 
and extrapolations of lower energy data.  
A key observable is the charged particle multiplicity, for which
data and three predictions are shown in the left panel of Fig.~\ref{fig:lhc}, 
from Ref.~\cite{Steinberg:2008zz}.
The coverage of ATLAS in the $\eta^{\prime}$ variable, relevant for testing
extended longitudinal scaling, is shown in the right panel.
Whether or not the data agree with current expectations,
there will surely be new insights on the nature of the perfect fluid.  
Of particular interest is the possibility that the system continues to thermalize
very early, something which will not occur if perturbative physics becomes more
important at higher energies.
The diversity of experiments and technologies will 
guarantee that the available phase space is fully covered for a variety of
observables, making the early days of LHC physics in both p+p and Pb+Pb 
particularly exciting.

\ack
The author would like to thank the PANIC 2008 organizers for the 
opportunity to speak in Eilat, in particular 
Itzhak Tserruya and Alexander Milov
who also invited several of us to an interesting 
post-conference workshop at the Weizmann Institute in Rehovot.

%
%
%

%
\end{document}